# MRFalign: Protein Homology Detection through Alignment of Markov Random Fields

Jianzhu Ma, Sheng Wang, Zhiyong Wang and Jinbo Xu[*]

Toyota Technological Institute at Chicago, 6045 Kenwood.Ave,
60637 Chicago, Illinois, USA
{majianzhu, wangsheng, j3xu}@ttic.edu

## Abstract

Sequence-based protein homology detection has been extensively studied and so far the most sensitive method is based upon comparison of protein sequence profiles, which are derived from multiple sequence alignment (MSA) of sequence homologs in a protein family. A sequence profile is usually represented as a position-specific scoring matrix (PSSM) or an HMM (Hidden Markov Model) and accordingly PSSM-PSSM or HMM-HMM comparison is used for homolog detection. This paper presents a new homology detection method MRFalign, consisting of three key components: 1) a Markov Random Fields (MRF) representation of a protein family; 2) a scoring function measuring similarity of two MRFs; and 3) an efficient ADMM (Alternating Direction Method of Multipliers) algorithm aligning two MRFs. Compared to HMM that can only model very short-range residue correlation, MRFs can model long-range residue interaction pattern and thus, encode information for the global 3D structure of a protein family. Consequently, MRF-MRF comparison for remote homology detection shall be much more sensitive than HMM-HMM or PSSM-PSSM comparison. Experiments confirm that MRFalign outperforms several popular HMM or PSSM-based methods in terms of both alignment accuracy and remote homology detection and that MRFalign works particularly well for mainly beta proteins. For example, tested on the benchmark SCOP40 (8353 proteins) for homology detection, PSSM-PSSM and HMM-HMM succeed on 48% and 52% of proteins, respectively, at superfamily level, and on 15% and 27% of proteins, respectively, at fold level. In contrast, MRFalign succeeds on 57.3% and 42.5% of proteins at superfamily and fold level, respectively. This study implies that long-range residue interaction patterns are very helpful for sequence-based homology detection. The software is available for download at http://raptorx.uchicago.edu/download/.

## Author Summary

Sequence-based protein homology detection has been extensively studied, but it remains very challenging for remote homologs with divergent sequences. So far the most sensitive methods employ HMM-HMM comparison, which models a protein family using HMM (Hidden Markov Model) and then detects homologs using HMM-HMM alignment. HMM cannot model long-range residue interaction patterns and thus, carries very little information regarding the global 3D structure of a protein family. As such, HMM comparison is not sensitive enough for distantly-related homologs. In this paper, we present an MRF-MRF comparison method for homology detection. In particular, we model a protein family using Markov Random Fields (MRF) and then detect homologs by MRF-MRF alignment. Compared to HMM, MRFs are able to model long-range residue interaction pattern and thus, contains information for the overall 3D structure of a protein family. Consequently, MRF-MRF comparison is much more sensitive than HMM-HMM comparison. To implement MRF-MRF comparison, we have developed a new scoring function to measure the similarity of two MRFs and also an efficient ADMM algorithm to optimize the scoring function. Experiments confirm that MRF-MRF comparison indeed outperforms HMM-HMM comparison in terms of both alignment accuracy and remote homology detection, especially for mainly beta proteins.

[*]To whom correspondence should be addressed

# 1   Introduction

Sequence-based protein alignment and homology detection has been extensively studied and widely applied to many biological problems such as homology modeling [1-4], phylogeny inference [5-7] and protein function prediction [8-10]. Although extensively studied, remote homology detection still remains very challenging, especially for homologs with divergent sequences. So far the most sensitive method for homology detection is based upon comparison of protein sequence profiles, which are usually derived from multiple sequence alignment (MSA) of sequence homologs in a protein family. That is, instead of aligning two primary sequences, homologs can be detected by aligning protein sequence profiles. To facilitate comparison and alignment, an MSA is usually represented as a position-specific scoring matrix (PSSM) [11] or an HMM (Hidden Markov Model) [12, 13]. HMM is more sensitive than PSSM because 1) HMM contains position-specific gap information; and 2) HMM also takes into account correlation among sequentially adjacent residues. Sequence signature libraries [14] and intermediate sequence based methods [15, 16] are also developed to make use of evolutionary information of a protein. All these methods are sensitive to close homologs, but not good enough for remote homologs. The main issue of existing profile-based methods lies in that they make use of only position-specific amino acid mutation patterns and very short-range residue correlation, but not long-range residue interaction. However, remote homologs may have very divergent sequences and are only similar at the level of (long-range) residue interaction pattern, which is not encoded in current popular PSSM or HMM models.

To significantly advance homology detection, this paper presents a Markov Random Fields (MRFs) modeling of a multiple sequence alignment (MSA). Compared to HMM, MRFs can model long-range residue interactions and thus, encodes information for the global 3D structure of a protein family. In particular, MRF is a graphical model encoding a probability distribution over the MSA by a graph and a set of preset statistical functions. A node in the MRF corresponds to one column in the MSA and one edge specifies correlation between two columns. Each node is associated with a function describing position-specific amino acid mutation pattern. Similarly, each edge is associated with a function describing correlated mutation statistics between two columns. With MRF representation, alignment of two proteins or protein families becomes that of two MRFs. To align two MRFs, a scoring function or alignment potential is needed to measure the similarity of two MRFs. We use a scoring function consists of both node alignment potential and edge alignment potential, which measure the node (i.e., amino acid) similarity and edge (i.e., interaction pattern) similarity, respectively.

It is computationally challenging to optimize a scoring function containing edge alignment potential. To deal with this, we formulate the MRF-MRF alignment problem as an integer programming problem and then develop an ADMM (Alternative Direction Method of Multipliers) algorithm to solve it efficiently to a suboptimal solution. ADMM divides the MRF alignment problem into two tractable sub-problems and then iteratively solve them until they reach consistent solutions.

Experiments show that our MRF-MRF alignment method, denoted as MRFalign, can generate more accurate alignments and is also much more sensitive than others in detecting remote homologs. MRFalign works particularly well on mainly-beta proteins.

**Related work**. Cowen has developed a program SMURFLite for fold recognition based upon the MRF representation of a protein family [17]. Nevertheless, our MRFalign method is significantly different from SMURFLite in a couple of aspects: 1) SMURLite builds an MRF based upon multiple structure alignment instead of multiple sequence alignment (MSA). As such, it cannot apply to sequence-based homology detection in the absence of native structures. In contrast, our method builds MRFs purely based upon MSA and thus, applies to sequence-based protein alignment and homology detection; and 2) SMURLite can only align a single primary sequence to an MRF, while our method aligns two MRFs to yield higher sensitivity. This difference requires us to develop totally new methods to build MRFs from MSA, measure similarity of two MRFs, and optimize the MRF-MRF alignment potential.

Quite a few PSSM-based profile comparison methods for homology detection have been developed, including [11, 18-23]. Some studies such as [20] also combine phylogeny information with PSSM-based profile comparison. Homology detection can also be done without aligning proteins. For example, we can represent a protein sequence or profile as a feature vector and then search for homologs by comparing feature vectors. Early methods such as [24] usually conduct straightforward comparison of feature vectors, but are not very sensitive [25]. Improvement in these alignment-free methods results from the application of discriminative learning approaches such as SVM–Fisher [26], SVM-pairwise [27], SVM with the spectrum kernel [28] and SVM with the mismatch kernel [29]. These SVM-based methods are reported to outperform the simple feature comparison methods. Comparing to alignment-based homology detection, alignment-free methods are usually faster but less sensitive.

## 2 Results

**Training and validation data.** To train the node alignment potential, we constructed the training and validation data from SCOP70. The sequence identity of all the training and validation protein pairs is uniformly distributed between 20% and 70%. Further, two proteins in any pair are similar at superfamily or fold level. In total we use a set of 1400 protein pairs as the training and validation data, which covers 458 SCOP folds [30]. Five-fold cross validation is used to choose the hyper-parameter in our machine learning model. In particular, every time we choose 1000 out of the 1400 protein pairs as the training data and the remaining 400 pairs as the validation data such that there is no fold-level redundancy between the training and validation data. A training or validation protein has less than 400 residues and contains less than 10% of residues without 3D coordinates. The reference alignment for a protein pair is generated by a structure alignment tool DeepAlign [31]. Each reference alignment has fewer than 50 gap positions in the middle and the number of terminal gaps is less than 20% of the alignment length.

**Test data.** The data used to test alignment accuracy has no fold-level overlap with the training and validation data. In particular, we use the following three datasets to test the alignment accuracy, which are subsets of the test data used in [4] to benchmark protein modeling methods.

1. Set3.6K: a set of 3617 non-redundant protein pairs. Two proteins in a pair share <40% sequence identity and have small length difference. By "non-redundant" we mean that in any two protein pairs, there are at least two proteins (one from each pair) sharing less than 25% sequence identity.

2. Set2.6K: a set of 2633 non-redundant protein pairs. Two proteins in a pair share <25% sequence identity and have length difference larger than 30%. This set is mainly used to test the performance of one method in handling with domain boundary.

3. Set60K: a very large set of 60929 protein pairs, in most of which two proteins share less than 40% sequence identity. Meanwhile, 846, 40902, and 19181 pairs are similar at the SCOP family, superfamily and fold level, respectively, and 151, 2691 and 2218 pairs consist of only all-beta proteins, respectively.

We use the following benchmarks to test remote homology detection success rate.

4. SCOP20, SCOP40 and SCOP80, which are used by Söding group to study context-specific mutation score [32]. They are constructed by filtering the SCOP database with a maximum sequence identity of 20%, 40% and 80%, respectively. In total they have 4884, 7088, and 9867 proteins, respectively, and 1281, 1806, and 2734 beta proteins, respectively.

We run PSI-BLAST with 5 iterations to detect sequence homologs and generate MSAs for the first three datasets. The MSA files for the three SCOP benchmarks are downloaded from the HHpred website (ftp://toolkit.genzentrum.lmu.de/pub/ ). Pseudocounts are used in building sequence profiles. Real secondary structure information is not used since this paper focuses on sequence-based homology detection.

**Programs to compare.** To evaluate alignment accuracy, we compare our method, denoted as MRFalign, with sequence-HMM alignment method HMMER [12] and HMM-HMM alignment method HHalign [13]. HHMER is run with a default E-value threshold (10.0). HHalign is run with the option "-mact 0.1". To evaluate the performance of homology detection, we compare MRFalign, with FFAS [11] (PSSM-PSSM comparison), hmmscan (sequence-HMM comparison) and HHsearch and HHblits [33] (HMM-HMM comparison). HHsearch and hmmscan use HHalign and HMMER, respectively, for protein alignment.

**Evaluation criteria.** Three performance metrics are used including reference-dependent alignment precision, alignment recall and homology detection success rate. Alignment precision is defined as the fraction of aligned positions that are correctly aligned. Alignment recall is the fraction of alignable residues that are correctly aligned. Reference alignments are used to judge if one residue is correctly aligned or alignable. To reduce bias, we use three very different structure alignment tools to generate reference alignments, including TM-align [34], Matt [35], and DeepAlign [31].

### Reference-dependent alignment recall

As shown in Tables 1 and 2, our method MRFalign exceeds all the others regardless of the reference alignments on both dataset Set3.6K and Set2.6K. MRFalign outperforms HHalign by ~10% on both datasets, and HHMER by ~23% and ~24%, respectively. If 4-position off the exact match is allowed in calculating alignment recall, MRFalign outperforms HHalign by ~11% on both datasets, and HHMER by ~25% and ~33%, respectively.

**(Table 1 should be placed here)**
**(Table 2 should be placed here)**

On the very large set Set60K, as shown in Table 3, our method outperforms the other two in each SCOP classification regardless of the reference alignments used. MRFalign is only slightly better than HHalign at the family level, which is not surprising since it is easy to align two closely-related proteins. At the superfamily level, our method outperforms HHalign and HMMER by ~6% and ~18%, respectively. At the fold level, our method outperforms HHalign and HHMER by ~7% and ~14%, respectively.

**(Table 3 should be placed here)**

**Alignment recall for beta proteins.** Our method outperforms HHalign and HMMER by ~3% and ~12%, respectively, at the family level; ~7% and ~19%, respectively, at the superfamily level; and ~10% and ~16%, respectively, at the fold level, regardless of reference alignments.

### Reference-dependent alignment precision

As shown in Tables 4 and 5, our method MRFalign exceeds all the others regardless of the reference alignments on both data sets Set3.6K and Set2.6K. MRFalign outperforms HHalign by ~8% and ~5%, respectively, and HMMER by ~15% and ~13%, respectively. If 4-position off the exact match is allowed in calculating alignment precision, MRFalign outperforms HHalign by ~8% and ~9%, and HMMER by ~14% and ~18% on Set3.6K and Set2.6K, respectively.

**(Table 4 should be placed here)**
**(Table 5 should be placed here)**

On the very large set Set60K, as shown in Table 6, our method outperforms the other two in each SCOP classification regardless of the reference alignments used. At the family level, our method outperforms HHalign and HMMER by ~3% and ~4%, respectively. At the superfamily level, our method outperforms HHalign and HMMER by ~4% and ~5%, respectively. At the fold level, our method outperforms HHalign and HHMER by ~5% and ~8%, respectively.

**(Table 6 should be placed here)**

### Homology detection success rate

To evaluate homology detection rate, we employ three benchmarks SCOP20, SCOP40 and SCOP80 introduced in [32]. For each protein sequence in one benchmark, we treat it as a query, align it to all the other proteins in the same benchmark and then examine if those with the best alignment scores are similar to the query or not. We also conducted homology detection experiments using hmmscan, FFAS, HHsearch and HHblits with default options. The success rate is measured at the superfamily and fold levels, respectively. When evaluating the success rate at the superfamily (fold) level, we exclude those proteins similar to the query at least at the family (superfamily) level. For each query protein, we examine the top 1-, 5- and 10-ranked proteins, respectively.

As shown in Table 7, tested on SCOP20, SCOP40 and SCOP80 at the superfamily level, our method MRFalign succeeds on ~6%, ~4% and ~4% more query proteins than HHsearch, respectively, when only the first-ranked proteins are considered. As shown in Table 8, at the fold level, MRFalign succeeds on ~11%, ~11% and ~12% more proteins than HHsearch, respectively, when only the first-ranked proteins are evaluated. At the superfamily level, SCOP20 is more challenging than the other two benchmarks because it contains fewer proteins similar at this level. Nevertheless, at the fold level, SCOP80 is slightly more challenging than the other two benchmarks maybe because it contains many more irrelevant proteins and thus, the chance of ranking false positives at top is higher.

Similar to alignment accuracy, our method for homology detection also has a larger advantage on the beta proteins. In particular, as shown in Table 9, tested on SCOP20, SCOP40 and SCOP80 at the superfamily level, MRFalign succeeds on ~7%, ~5% and ~7% more proteins than HHsearch, respectively, when only the first-ranked proteins are evaluated. As shown in Table 10, at the fold level, MRFalign succeeds on ~13%, ~16% and ~17% more proteins than HHsearch, respectively, when only the first-ranked proteins are evaluated. Note that in this experiment, only the query proteins are mainly-beta proteins, the subject proteins can be of any types. If we restrict the subject proteins to only beta proteins, the success rate increases further due to the reduction of false positives.

**(Table 7 should be placed here)**

**(Table 8 should be placed here)**

**(Table 9 should be placed here)**

**(Table 10 should be placed here)**

**Contribution of edge alignment potential and mutual information**
To evaluate the contribution of our edge alignment potential, we calculate the alignment recall improvement resulting from using edge alignment potential on two benchmarks Set3.6K and Set2.6K. As shown in Table 11, our edge alignment potential can improve alignment recall by 3.4% and 3.7%, respectively. When mutual information is used, we can further improve alignment recall by 1.1% and 1.9% on these two sets, respectively. Mutual information is mainly useful for proteins with many sequence homologs since it is close to 0 when there are few sequence homologs. As shown in Table 11, if only those proteins with at least 256 non-redundant sequence homologs are considered, the improvement resulting from mutual information is ~3%.

**(Table 11 should be placed here)**

**Running time**
Figure 1 shows the running time of MRFalign with respect to protein length. As a control, we also show the running time of the Viterbi algorithm, which is used by our ADMM algorithm to generate alignment at each iteration. As shown in this figure, MRFalign is no more than 10 times slower than the Viterbi algorithm. To speed up homology detection, we first use the Viterbi algorithm to perform an initial search without considering edge alignment potential and keep only top 200 proteins, which are then subject to realignment and rerank by our MRFalign method. Therefore, although MRFalign may be very slow compared to the Viterbi algorithm, empirically we can do homology search only slightly slower than the Viterbi algorithm.

**(Figure 1 should be placed here)**

**Is our MRFalign method overtrained?**

We conducted two experiments to show that our MRFalign is not overtrained. In the first experiment, we used 36 CASP10 hard targets as the test data. Our training set was built before CASP10 started, so there is no redundancy between the CASP10 hard targets and our training data. Using MRFalign and HHpred, respectively, we search each of these 36 test targets against PDB25 to find the best match. Since PDB25 does not contain proteins very similar to many of the test targets, we built a 3D model using MODELLER from the alignment between a test target and its best match and then measure the quality of the model. As shown in Figure 2, MRFalign can yield much better 3D models than HHsearch for most of the targets. This implies that our method can generalize well to the test data not similar to the training data.

In the second experiment, we divide the proteins in SCOP40 into three subsets according their similarity with all the training data. We measure the similarity of one test protein with all the training data by its best BLAST E-value. We used two values 1e-2 and 1e-35 as the E-value cutoff so that the three subsets have roughly the same size. As shown in Table 12, the advantage of our method in remote homology detection over HHpred is roughly same across the three subsets. Since HHpred is an unsupervised algorithm, this implies that the performance of our method is not correlated to the test-training similarity. Therefore, it is unlikely that our method is overfit by the training data.

(Figure 2 should be placed here)

(Table 12 should be placed here)

## 3 Discussion

This paper has presented a new method for sequence-based protein homology detection that compares two protein sequences or families through alignment of two Markov Random Fields (MRFs), which model the multiple sequence alignment (MSA) of a protein family using an undirected general graph in a probabilistic way. The MRF representation is better than the extensively-used PSSM and HMM representations in that the former can capture long-range residue interaction pattern, which reflects the overall 3D structure of a protein family. As such, MRF comparison is much more sensitive than HMM comparison in detecting remote homologs. This is validated by our large-scale experimental tests showing that MRF-MRF comparison can greatly improve alignment accuracy and remote homology detection over currently popular sequence-HMM, PSSM-PSSM, and HMM-HMM comparison methods. Our method also has a larger advantage over the others on mainly-beta proteins.

We build our MRF model of a protein family based upon multiple sequence alignment (MSA) in the absence of native structures. The accuracy of the MRF model depends on the accuracy of an MSA. Currently we rely on the MSA generated by PSI-BLAST. In the future, we may explore better alignment methods for MSA building or even utilize solved structures of one or two protein sequences to improve MSA. The accuracy of the MRF model parameter usually increases with respect to the number of non-redundant sequence homologs in the MSA. Along with more and more protein sequences are generated by a variety of sequencing projects, we shall be able to build accurate MRFs for more and more protein families and thus, detect their homologous relationship more accurately.

An accurate scoring function is essential to MRF-MRF comparison. Many different methods can be used to measure node and edge similarity of two MRFs, just like many different scoring functions can be used to measure the similarity of two PSSMs or HMMs. This paper presents only one of them. In the future we may explore more possibilities. It is computationally intractable to find the best alignment between two MRFs when edge similarity is taken into consideration. This paper presents an ADMM algorithm that can efficiently solve the MRF-MRF alignment problem to suboptimal. However, this algorithm currently is about 10 times slower than the Viterbi algorithm for PSSM-PSSM alignment. Further tuning of this ADMM algorithm is needed for very large-scale homology detection.

## 4 Methods

### 4.1 Modeling Multiple Sequence Alignment (MSA) Using Markov Random Fields (MRF)

Given a protein primary sequence, we run PSI-BLAST [37] with 5 iterations and E-value cutoff 0.001 to find its sequence homologs. PSI-BLAST also generates an MSA of the sequence homologs. Let $X_i$ be a finite discrete random variable representing the amino acid at column $i$ in the MSA, taking values from 1 to 21, corresponding to 20 amino acids and gap. Then we can use a multivariate random variable $X = \{X_1, X_2, \ldots, X_N\}$, where $N$ is the number of columns, to model the MSA. We use an MRF to define the probability distribution of $X$. MRF is an undirected graph that can be used to model a set of correlated random variables. As shown in Fig. 3, an MRF node represents one column in the MSA and an edge represents the correlation between two columns $i$ and $k$ when $|i - k| \geq 6$. We ignore very short-range correlation (i.e., $|i - k| < 6$) since it is not very informative. The MRF consists of two types of functions: $\phi(X_i)$ and $\psi(X_i, X_k)$, where $\phi(X_i)$ is an amino acid preference function for node $i$ and $\psi(X_i, X_k)$ is a pairwise amino acid preference function for edge $(i, k)$ that reflects interaction between two nodes. Then, the probability of observing a particular protein sequence $X$ can be calculated as follows.

$$P(X) = \frac{1}{Z} \prod_i \phi(X_i) \prod_{(i,k)} \psi(X_i, X_k), \tag{1}$$

where $Z$ is the normalization factor.

**(Figure 3 should be placed here)**

We use two kinds of information in MRFs for their alignment. One is the occurring probability of 20 amino acids and gap at each node (i.e., each column in MSA), which can also be interpreted as the marginal probability at each node. The other is the correlation between two nodes, which can be interpreted as interaction strength of two MSA columns and calculated by several different ways. For example, we can use a contact prediction program such as PSICOV [38] and PhyCMAP [39] for this purpose. PSICOV assumes that $P(X)$ is a Gaussian distribution function and calculates the correlation between two columns by inverse covariance matrix. PhyCMAP takes sequence information (including mutual information) as input and predicts the probability of two residues forming a contact, which can be used to indicate the interaction strength of two columns. However, it takes time to run these programs, in current implementation we calculate the mutual information (MI) and its power series of two columns as interaction strength. That is, we use MI, $MI^2$, ..., $MI^{11}$ to quantify all the pairwise interaction strength where MI is the mutual information matrix. The MI power series are much more informative than the MI alone, as tested in our contact prediction program PhyCMAP.

### 4.2 Scoring function for the alignment of two Markov Random Fields (MRFs)

Our scoring function for MRF-MRF alignment is a linear combination of node alignment potential and edge alignment potential with equal weight. Let $T$ and $S$ denote two MRFs for the two proteins under consideration. There are three possible alignment states $M$, $I_t$ and $I_s$ where $M$ represents two nodes being aligned, $I_t$ denotes an insertion in $T$ (i.e., one node in $T$ is not aligned), and $I_s$ denotes an insertion in $S$ (i.e., one node in $S$ is not aligned). As shown in Fig. 4, each alignment can be represented as a path in an alignment matrix, in which each vertex can be exactly determined by its position in the matrix and its state. For example, the first vertex in the path can be written as (0, 0, dummy), the 2nd vertex as $(1, 1, M)$ and the 3rd vertex as $(2, 1, I_s)$. Therefore, we can write an alignment as a set of triples, each of which has a form like $(i, j, u)$ where $(i, j)$ represents the position and $u$ the state.

**(Figure 4 should be placed here)**

**Node alignment potential.** Given an alignment path, its node alignment potential is the accumulative potential of all the vertices in the path. We use a Conditional Neural Fields (CNF) [40] method, which is very similar to what is described in the protein threading paper [41], to estimate the occurring probability of an alignment and then derive node alignment potential from this CNF. Briefly speaking, we estimate the occurring probability of an alignment $A$ between $T$ and $S$ as follows.

$$P(A|T,S) = e^{\sum_{(i,j,u)\in A} E_u(T_i, S_j)}/Z(T,S), \qquad (2)$$

where $Z(T,S)$ is a normalization factor summarizing all the possible alignments between $T$ and $S$, and $E_u(T_i, S_j)$ is a neural network with one hidden layer that calculates the log-likelihood of a vertex $(i,j,u)$ in the alignment path, where $i$ is a node in $T$, $j$ a node in $S$, and $u$ a state. When $u$ is a match state, $E_u$ takes as input the sequence profile context of two nodes $i$ and $j$, denoted as $T_i$ and $S_j$, respectively, and yields the log-likelihood of these two nodes being matched. When $u$ is an insertion state, it takes as input at the sequence profile context of one node and yields the log-likelihood of this node being an insertion. The sequence profile context of node $i$ is a $21 \times (2w + 1)$ matrix where $w = 5$, consisting of the marginal probability of 20 amino acids and gap at $2w + 1$ nodes indexed by $i - w$, $i - w + 1$,…, $i$, $i + 1$, …, $i + w$. In case that one column does not exist (when $i \leq w$ or $i + w > N$), zero is used. We train the parameters in $E_u$ by maximizing the occurring probability of a set of reference alignments, which are generated by a structure alignment tool DeepAlign [31]. That is, we optimize the model parameters so that the structure alignment of one training protein pair has the largest probability among all possible alignments. A $L_2$-norm regularization factor, which is determined by 5-fold cross validation, is used to restrict the search space of model parameters to avoid over-fitting. See the paper [41] for more technical details.

Let $\theta_{i,j}^u$ denote the potential of a vertex $(i,j,u)$ in the alignment path. We calculate $\theta_{i,j}^u$ from $E_u$ as follows.

$$\theta_{i,j}^u = E_u(T_i, S_j) - Exp(E_u), \qquad (3)$$

where $Exp(E_u)$ is the expected value of $E_u$. It is used to offset the effect of the background, which is the log-likelihood yielded by $E_u$ for any randomly-chosen node pairs (or nodes). Once $E_u$ is determined, we can approximate its expected value by sampling. That is, we sample ten thousands of node pairs (or nodes) from the training data, feed their sequence information into $E_u$ and then calculate the average output of $E_u$ as its expected value.

**Edge alignment potential.** It calculates the similarity of two edges, one from each MRF, based upon the interaction strength of two ends in one edge. We can derive interaction strength from the parameters of the MRF model, but it is hard to validate if this interaction strength (or mutual information) is accurate or not even in the presence of native structures since we cannot directly measure interaction strength in a protein. Here we use inter-residue Euclidean distance, which can be measured more easily, to reflect interaction strength of two residues. Later in this section we will describe how to derive the distance probability distribution from the information (e.g., interaction strength) encoded in MRFs. Let $d_{ik}^T$ denote the Euclidean distance between two residues at $i$ and $k$ and $d_{jl}^S$ is defined similarly. Note that $d_{ik}^T$ and $d_{jl}^S$ are unknown since this paper studies sequence-based homology detection in the absence of native structures. Let $\theta_{i,k,j,l}$ denote the alignment potential between edge $(i,k)$ in $T$ and edge $(j,l)$ in $S$. We calculate $\theta_{i,k,j,l}$ as follows.

$$\theta_{i,k,j,l} = \sum_{d_{ik}^T, d_{jl}^S} p(d_{ik}^T|c_i, c_k, m_{ik}) p(d_{jl}^S|c_j, c_l, m_{jl}) \log \frac{p(d_{ik}^T, d_{jl}^S)}{P_{ref}(d_{ik}^T) P_{ref}(d_{jl}^S)}, \qquad (4)$$

where $p(d_{ik}^T|c_i, c_k, m_{ik})$ is the probability of two nodes $i$ and $k$ in $T$ interacting at distance $d_{ik}^T$; $p(d_{jl}^S|c_j, c_l, m_{jl})$ is the probability of two nodes $j$ and $l$ in $S$ interacting at distance $d_{jl}^S$; $p(d_{ik}^T, d_{jl}^S)$ is the probability of one distance $d_{ik}^T$ being aligned to another distance $d_{jl}^S$ in reference alignments; and $P_{ref}(d_{ik}^T)$ $(P_{ref}(d_{jl}^S))$ is the background probability of observing $d_{ik}^T$ $(d_{jl}^S)$ in a protein structure.

Now we explain how to calculate each term in Eq. (4). $P_{ref}(d_{ik}^T)$ $(P_{ref}(d_{jl}^S))$ can be calculated by simple counting on a set of non-redundant protein structures, e.g., PDB25. Similar to $P_{ref}(d_{ik}^T)$, $P(d_{ik}^T, d_{jl}^S)$ can also be calculated by simple counting on a set of non-redundant reference alignments. That is, we randomly choose a set of protein pairs such that two proteins in each pair are similar at least at the fold level. Then we generate their reference alignment (i.e., structure alignments) using a structure alignment tool DeepAlign [31] and finally do simple counting to estimate $p(d_{ik}^T, d_{jl}^S)$. In order to use simple counting, we discretize inter-residue distance into 12 intervals: <4Å, 4-5Å, 5-6Å, …, 14-15Å, and >15Å.

In Eq. (4), $p(d_{ik}^T|c_i, c_k, m_{ik})$ is the probability of two nodes $i$ and $k$ interacting at distance $d_{ik}^T$, conditioned on information derived from the MRF. $p(d_{jl}^S|c_j, c_l, m_{jl})$ is defined similarly. Meanwhile, $c_i$ and $c_k$ are the sequence profile contexts of two nodes $i$ and $k$, respectively, and $m_{ik}$ represents the mutual information and its power series (or interaction strength) between these two nodes. The sequence profile context of node $i$ is a $21 \times (2w + 1)$ matrix where $w = 7$, consisting of the occurring probability of 20 amino acids and gap at $2w + 1$ nodes indexed by $i - w$, $i - w + 1$,…, $i$, $i + 1$, …, $i + w$. In case that one column does not exist (when $i \leq w$ or $i + w > N$), zero is used. We predict $p(d_{ik}^T|c_i, c_k, m_{ik})$ using a probabilistic neural network (PNN) implemented in our context-specific distance-dependent statistical potential package EPAD [36]. EPAD takes as input sequence contexts and mutual information and then yields inter-residue distance probability distribution. Compared to contact information, here we use interaction at a given distance to obtain a higher-resolution description of the residue interaction pattern. Therefore, our scoring function contains more information and thus, may yield better alignment accuracy and homology detection rate.

### 4.3 Aligning two MRFs by ADMM (Alternating Direction Method of Multipliers)

As mentioned before, an alignment can be represented as a path in the alignment matrix, which encodes an exponential number of paths. We can use a set of $3N_1N_2$ binary variables $\{z_{i,j}^u \mid 1 \leq i \leq N_1, 1 \leq j \leq N_2, u \in \{M, I_s, I_t\}\}$ to indicate which path is chosen, where $N_1$ and $N_2$ are the lengths of the two MSAs, $(i, j)$ is an entry in the alignment matrix and $u$ is the associated state. $z_{i,j}^u$ is equal to 1 if the alignment path passes $(i, j)$ with state $u$. Therefore, the problem of finding the best alignment between two MRFs can be formulated as the following quadratic optimization problem.

$$(P1) \quad \max_z \sum_{i,j,u} \theta_{i,j}^u z_{i,j}^u + \frac{1}{L}\sum_{i,j,k,l,u,v} \theta_{i,j,k,l}^{uv} z_{i,j}^u z_{k,l}^v, \quad (5)$$

where $\theta_{i,j}^u$ and $\theta_{i,j,k,l}^{uv}$ are node and edge alignment potentials as described in previous section. Meanwhile, $\theta_{i,j,k,l}^{uv}$ is equal to 0 if either $u$ or $v$ is not a match state. $L$ is the alignment length and $1/L$ is used to make the accumulative node and edge potential have similar scale. Note that $L$ is unknown and we will describe how to determine it later in this section. Finally, the solution of P1 shall be subject to the constraint that all those $z_{i,j}^u$ with value 1 shall form a valid alignment path. This constraint shall also be enforced to all the optimization problems described in this section.

It is computationally intractable to find the optimal solution of P1. Below we present an ADMM (Alternating Direction Method of Multipliers) method that can efficiently solve this problem to suboptimal. See [42] for a tutorial of the ADMM method. To use ADMM, we rewrite P1 as follows by making a copy of $z$ to $y$, but without changing the solution space.

$$(P2) \quad \max_{z,y} \sum_{i,j,u} \theta_{i,j}^u z_{i,j}^u + \frac{1}{L}\sum_{i,j,k,l,u,v} \theta_{i,j,k,l}^{uv} z_{i,j}^u y_{k,l}^v \quad (6)$$

$$s.t. \; \forall k, l, v, \quad z_{k,l}^v = y_{k,l}^v$$

Problem P2 can be augmented by adding a term to penalize the difference between $z$ and $y$.

$$(\text{P3}) \quad \max_{z,y} \sum_{i,j,u} \theta_{i,j}^u z_{i,j}^u + \frac{1}{L}\sum_{i,j,k,l,u,v} \theta_{i,j,k,l}^{uv} z_{i,j}^u y_{k,l}^v - \frac{\rho}{2}\sum_{i,j,u}(z_{i,j}^u - y_{i,j}^u)^2 \quad (7)$$

$$s.t. \ \forall i,j,u, \quad z_{i,j}^u = y_{i,j}^u$$

P3 is equivalent to P2 and P1, but converges faster due to the penalty term. Here $\rho$ is a hyper-parameter influencing the convergence rate of the algorithm. Some heuristics algorithms were proposed for choosing $\rho$ at each iteration, such as [43, 44]. Empirically, setting $\rho$ to a constant (=0.5) enables our algorithm to converge within 10 iterations for most protein pairs.

Adding the constraint $z_{i,j}^u = y_{i,j}^u$ using a Lagrange multiplier $\lambda$ to Eq. (7), we have the following Lagrangian dual problem:

$$(\text{P4}) \ \min_\lambda \max_{z,y} \sum_{i,j,u} \theta_{i,j}^u z_{i,j}^u + \frac{1}{L}\sum_{i,j,k,l,u,v} \theta_{i,j,k,l}^{uv} z_{i,j}^u y_{k,l}^v + \sum_{i,j,u} \lambda_{i,j}^u (z_{i,j}^u - y_{i,j}^u) - \frac{\rho}{2}\sum_{i,j,u}(z_{i,j}^u - y_{i,j}^u)^2 \quad (8)$$

It is easy to prove that P3 is upper bounded by P4. Now we will solve P4 and use its solution to approximate P3 and thus, P1.

Since both $z$ and $y$ are binary variables, the last term in Eq. (8) can be expanded as follows.

$$\frac{\rho}{2}\sum_{i,j,u}(z_{i,j}^u - y_{i,j}^u)^2 = \frac{\rho}{2}\sum_{i,j,u}(z_{i,j}^u + y_{i,j}^u - 2z_{i,j}^u y_{i,j}^u) \quad (9)$$

For a fixed $\lambda$, we can split P4 into the following two sub-problems.

$$(\text{SP1}) \quad y^* = argmax\{\sum_{k,l,v} y_{k,l}^v C_{k,l}^v\}, \quad (10)$$

$$\text{where } C_{k,l}^v = \frac{1}{L}\sum_{i,j,u} \theta_{i,j,k,l}^{uv} z_{i,j}^u - \lambda_{k,l}^v - \frac{\rho}{2}(1 - 2z_{k,l}^v)$$

$$(\text{SP2}) \quad z^* = argmax\{\sum_{i,j,u} z_{i,j}^u D_{i,j}^u\}, \quad (11)$$

$$\text{where } D_{i,j}^u = \theta_{i,j}^u + \sum_{k,l,v}\frac{1}{L}\theta_{i,j,k,l}^{uv} y_{k,l}^{v*} + \lambda_{i,j}^u - \frac{\rho}{2}(1 - y_{i,j}^{u*})$$

The sub-problem SP1 optimizes the objective function with respect to $y$ while fixing $z$, and the sub-problem SP2 optimizes the objective function with respect to $z$ while fixing $y$. SP1 and SP2 do not contain any quadratic term, so they can be efficiently solved using the classical dynamic programming algorithm for sequence or HMM-HMM alignment.

In summary, we solve P4 using the following procedure.
1) Initialize $z$ by aligning the two MRFs without the edge alignment potential, which can be done by dynamic programming. Accordingly, initialize $L$ as the length of the initial alignment.
2) Solve (SP1) first and then (SP2) using dynamic programming, each generating a feasible alignment.
3) If the algorithm converges, i.e., the difference between $z$ and $y$ is very small or zero, stop here. Otherwise, we update the alignment length $L$ as the length of the alignment just generated and the Lagrange multiplier $\lambda$ using subgradient descent as in Eq. (12), and then go back to Step 2).

$$\lambda^{n+1} = \lambda^n - \rho(z^* - y^*) \quad (12)$$

Due to the quadratic penalty term in Eq. (7), this ADMM algorithm usually converges much faster and also yields better solutions than without this term. Empirically, it converges within 10 iterations for most protein pairs. See [42] for the convergence proof of a general ADMM algorithm.

# Figures

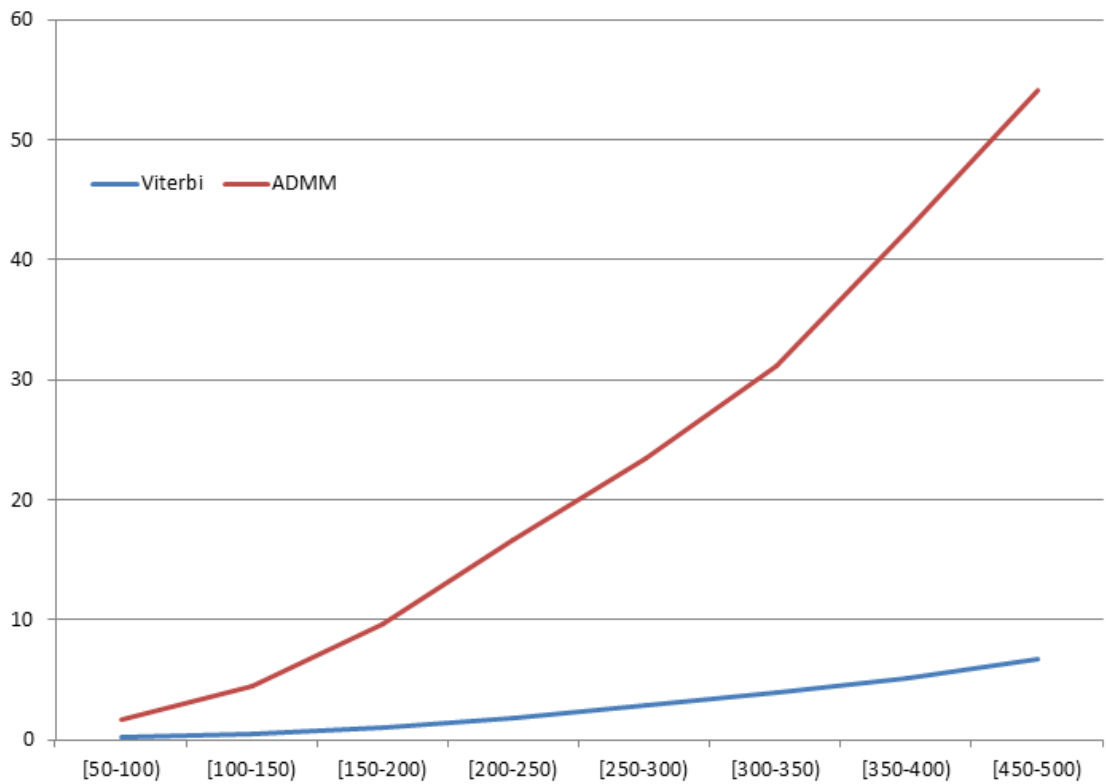

**Figure 1.** Running time of the Viterbi algorithm and our ADMM algorithm. The X-axis is the geometric mean of the two protein lengths in a protein pair. The Y-axis is the running time in seconds.

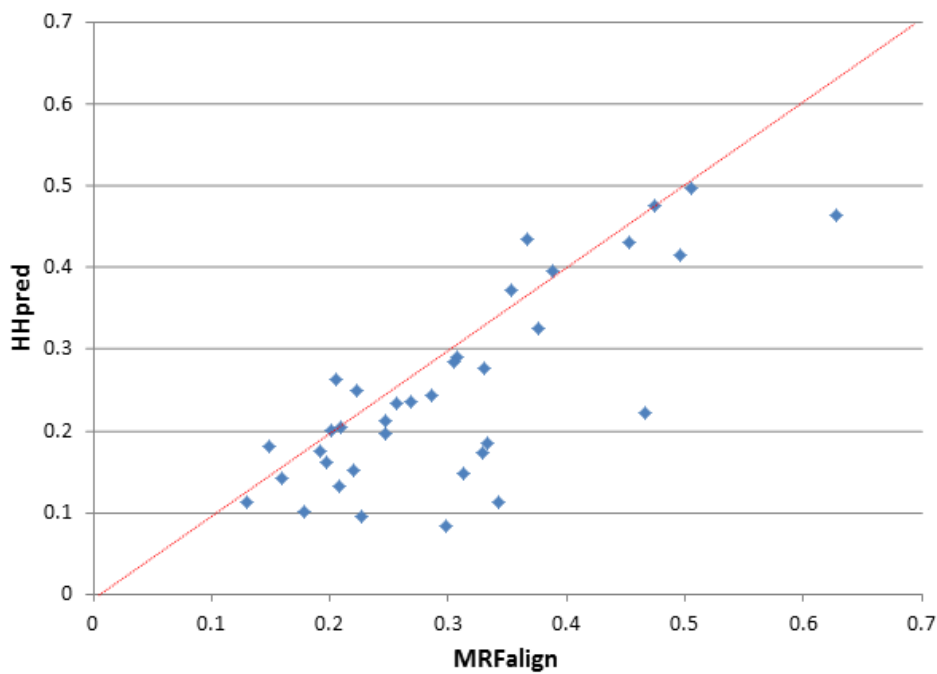

**Figure 2.** The model quality, measured by TM-score, of our method and HHpred for the 36 CASP10 hard targets. One point represents two models generated by our method (x-axis) and HHpred (y-axis).

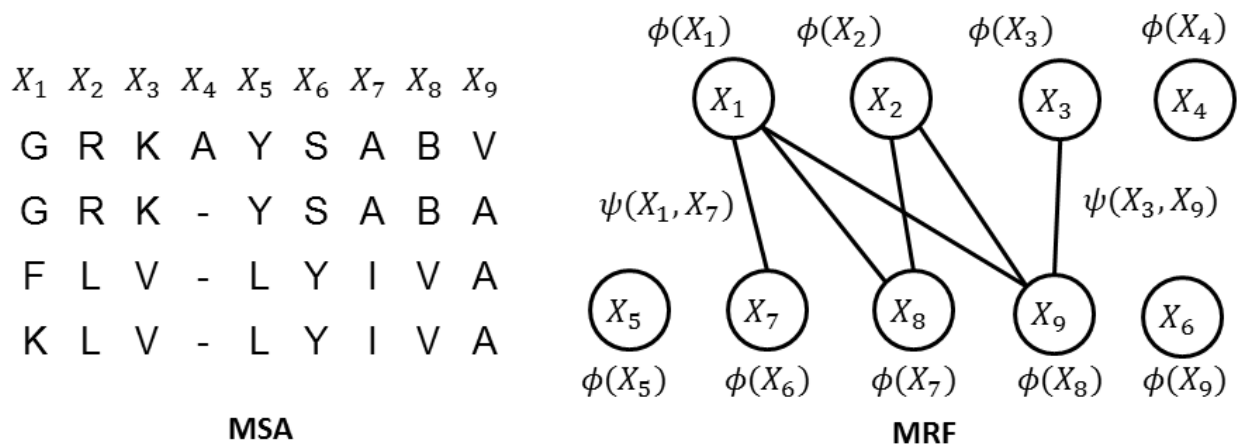

**Figure 3.** Model a multiple sequence alignment (left) by a Markov Random Fields (right).

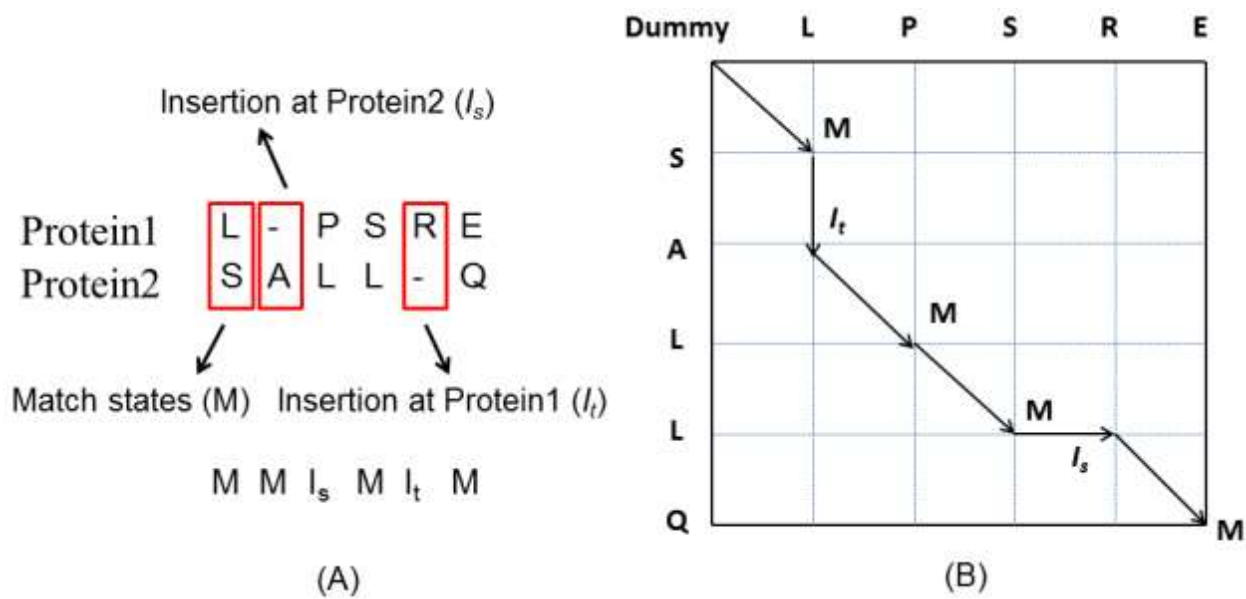

**Figure 4.** Representation of protein alignment. (A) Represented as a sequence of states. (B) Each alignment is a path in the alignment matrix.

# Tables

**Table 1.** Reference-dependent alignment recall on Set3.6K. Three structure alignment tools (TMalign, Matt and DeepAlign) are used to generate reference alignments. "4-offset" means that 4-position off the exact match is allowed. The bold indicates the best results.

|  | TMalign | | Matt | | DeepAlign | |
|---|---|---|---|---|---|---|
|  | Exact match | 4-offset | Exact match | 4-offset | Exact match | 4-offset |
| HMMER | 22.9% | 26.5% | 24.1% | 27.4% | 25.5% | 28.1% |
| HHalign | 36.3% | 39.1% | 37.0% | 42.1% | 38.4% | 42.8% |
| MRFalign | **47.4%** | **51.0%** | **47.5%** | **52.6%** | **49.2%** | **53.5%** |

**Table 2.** Reference-dependent alignment recall on Set2.6K. See Table 1 for explanation.

|  | TMalign | | Matt | | DeepAlign | |
|---|---|---|---|---|---|---|
|  | Exact match | 4-offset | Exact match | 4-offset | Exact match | 4-offset |
| HMMER | 36.5% | 42.6% | 38.6% | 44.0% | 40.4% | 45.0% |
| HHalign | 62.5% | 66.1% | 63.2% | 66.2% | 64.0% | 66.7% |
| MRFalign | **72.8%** | **76.2%** | **73.5%** | **76.7%** | **74.2%** | **77.8%** |

**Table 3.** Reference-dependent alignment recall (exact match) on the large benchmark Set60K. The protein pairs are divided into 3 groups based upon the SCOP classification. The bold indicates the best results.

| | TMalign | | | Matt | | | DeepAlign | | |
|---|---|---|---|---|---|---|---|---|---|
| | HMMER | HHalign | MRFalign | HMMER | HHalign | MRFalign | HMMER | HHalign | MRFalign |
| Family | 57.4% | 69.2% | **71.0%** | 59.1% | 70.5% | **74.5%** | 63.2% | 72.6% | **75.5%** |
| Superfamily | 31.2% | 42.0% | **48.1%** | 32.3% | 42.4% | **51.7%** | 32.8% | 49.4% | **55.6%** |
| Fold | 1.3% | 7.0% | **14.2%** | 1.6% | 8.0% | **15.5%** | 2.0% | 8.7% | **18.4%** |
| Family (beta) | 60.9% | 69.9% | **73.1%** | 64.0% | 75.1% | **78.4%** | 68.4% | 79.0% | **82.9%** |
| Superfamily (beta) | 35.0% | 47.2% | **52.1%** | 37.0% | 50.2% | **55.8%** | 39.1% | 52.9% | **60.7%** |
| Fold (beta) | 2.5% | 8.3% | **17.3%** | 3.0% | 9.1% | **17.1%** | 4.0% | 10.1% | **21.8%** |

**Table 4.** Reference-dependent alignment precision on Se3.6K. Three structure alignment tools (TMalign, Matt and DeepAlign) are used to generate reference alignments. "4-offset" means that 4-position off the exact match is allowed. The bold indicates the best results.

| | TMalign | | Matt | | DeepAlign | |
|---|---|---|---|---|---|---|
| | Exact match | 4-offset | Exact match | 4-offset | Exact match | 4-offset |
| HMMER | 29.3% | 34.1% | 29.6% | 34.7% | 31.5% | 35.6% |
| HHalign | 35.9% | 39.4% | 36.2% | 39.4% | 37.2% | 41.7% |
| MRFalign | **43.2%** | **47.4%** | **44.1%** | **48.5%** | **46.1%** | **50.4%** |

**Table 5.** Reference-dependent alignment precision on Set2.6K. See Table 4 for explanation.

| | TMalign | | Matt | | DeepAlign | |
|---|---|---|---|---|---|---|
| | Exact match | 4-offset | Exact match | 4-offset | Exact match | 4-offset |
| HMMER | 48.0% | 50.1% | 48.2% | 50.3% | 51.4% | 54.8% |
| HHalign | 57.1% | 59.9% | 57.3% | 60.0% | 58.3% | 61.4% |
| MRFalign | **62.5%** | **69.1%** | **62.7%** | **69.6%** | **63.2%** | **70.0%** |

**Table 6.** Reference-dependent alignment precision (exact match) on the large benchmark Set60K. The protein pairs are divided into 3 groups based upon the SCOP classification. The bold indicates the best results.

| | TMalign | | | Matt | | | DeepAlign | | |
|---|---|---|---|---|---|---|---|---|---|
| | HMMER | HHalign | MRFalign | HMMER | HHalign | MRFalign | HMMER | HHalign | MRFalign |
| Family | 63.1% | 63.9% | **67.3%** | 64.3% | 65.4% | **68.0%** | 68.4% | 69.2% | **71.4%** |
| Superfamily | 38.7% | 39.5% | **42.8%** | 40.5% | 41.3% | **44.9%** | 43.2% | 44.3% | **48.7%** |
| Fold | 4.2% | 7.4% | **11.5%** | 4.7% | 8.0% | **12.3%** | 5.4% | 8.2% | **14.5%** |
| Family (beta) | 66.4% | 65.8% | **69.5%** | 67.4% | 68.1% | **72.3%** | 70.8% | 72.4% | **77.9%** |
| Superfamily (beta) | 44.2% | 44.9% | **48.8%** | 45.4% | 46.2% | **49.4%** | 46.6% | 48.4% | **53.7%** |
| Fold (beta) | 6.1% | 9.3% | **14.1%** | 6.7% | 9.2% | **14.5%** | 7.9% | 8.6% | **17.8%** |

**Table 7.** Homology detection performance at the superfamily level

| | Scop20 | | | Scop40 | | | Scop80 | | |
|---|---|---|---|---|---|---|---|---|---|
| | Top1 | Top5 | Top10 | Top1 | Top5 | Top10 | Top1 | Top5 | Top10 |
| hmmscan | 35.2% | 36.5% | 36.5% | 40.2% | 41.7% | 41.8% | 43.9% | 45.2% | 45.3% |
| FFAS | 48.6% | 54.4% | 55.6% | 52.1% | 56.3% | 57.1% | 49.8% | 53.0% | 53.7% |
| HHsearch | 51.6% | 57.3% | 59.2% | 55.8% | 60.8% | 62.4% | 56.1% | 60.1% | 61.8% |
| HHblits | 51.9% | 56.3% | 57.5% | 56.0% | 59.8% | 60.9% | 59.2% | 62.5% | 63.3% |
| MRFalign | **58.2%** | **61.7%** | **63.4%** | **59.3%** | **63.6%** | **65.8%** | **60.4%** | **64.7%** | **66.1%** |

**Table 8.** Homology detection performance at the fold level

| | Scop20 | | | Scop40 | | | Scop80 | | |
|---|---|---|---|---|---|---|---|---|---|
| | Top1 | Top5 | Top10 | Top1 | Top5 | Top10 | Top1 | Top5 | Top10 |
| hmmscan | 5.2% | 6.1% | 6.1% | 6.2% | 6.9% | 6.9% | 5.9% | 6.5% | 6.6% |
| FFAS | 13.1% | 18.7% | 20.0% | 10.4% | 14.5% | 15.4% | 9.1% | 11.9% | 12.6% |

| | | | | | | | | |
|---|---|---|---|---|---|---|---|---|
| HHsearch | 16.3% | 24.7% | 28.6% | 17.6% | 25.3% | 29.1% | 15.4% | 21.9% | 25.0% |
| HHblits | 17.4% | 25.2% | 27.2% | 19.1% | 26.0% | 28.2% | 18.4% | 25.0% | 27.0% |
| MRFalign | **27.2%** | **36.8%** | **41.2%** | **28.3%** | **37.9%** | **42.4%** | **27.0%** | **38.1%** | **41.6%** |

**Table 9.** Homology detection performance for mainly beta proteins at the superfamily level

| | Scop20 | | | Scop40 | | | Scop80 | | |
|---|---|---|---|---|---|---|---|---|---|
| | Top1 | Top5 | Top10 | Top1 | Top5 | Top10 | Top1 | Top5 | Top10 |
| hmmscan | 29.1% | 29.4% | 29.4% | 34.7% | 35.1% | 35.1% | 43.7% | 44.0% | 44.1% |
| FFAS | 43.6% | 49.9% | 51.9% | 48.2% | 52.4% | 53.5% | 43.7% | 46.3% | 47.2% |
| HHsearch | 48.2% | 54.6% | 56.9% | 52.0% | 56.9% | 59.1% | 47.7% | 51.8% | 53.7% |
| HHblits | 47.5% | 52.1% | 53.7% | 51.4% | 54.8% | 56.6% | 52.9% | 54.6% | 57.8% |
| MRFalign | **55.4%** | **61.7%** | **65.9%** | **57.3%** | **63.5%** | **66.8%** | **54.2%** | **59.7%** | **64.2%** |

**Table 10.** Homology detection performance for mainly beta proteins at the fold level

| | Scop20 | | | Scop40 | | | Scop80 | | |
|---|---|---|---|---|---|---|---|---|---|
| | Top1 | Top5 | Top10 | Top1 | Top5 | Top10 | Top1 | Top5 | Top10 |
| hmmscan | 6.9% | 7.6% | 7.6% | 8.0% | 8.6% | 8.6% | 7.0% | 7.4% | 7.4% |
| FFAS | 22.7% | 30.1% | 31.8% | 15.2% | 20.4% | 21.7% | 11.8% | 15.3% | 16.1% |
| HHsearch | 24.4% | 34.7% | 38.8% | 26.8% | 37.7% | 41.6% | 19.1% | 26.8% | 29.5% |
| HHblits | 24.1% | 33.3% | 34.8% | 26.9% | 35.3% | 37.1% | 24.7% | 34.1% | 35.5% |
| MRFalign | **37.4%** | **55.0%** | **61.4%** | **42.5%** | **51.1%** | **54.6%** | **36.4%** | **48.0%** | **55.9%** |

**Table 11.** Contribution of edge alignment potential and mutual information, measured by alignment recall improvement on two benchmarks Set3.6K and Set2.6K. The structure alignments generated by DeepAlign are used as reference alignments.

| **Alignment recall for the whole test sets** | | | | |
|---|---|---|---|---|
| | Set3.6K | | Set2.6K | |
| | Exact Match | 4-position offset | Exact Match | 4-position offset |
| Only with node potential | 44.7% | 48.6% | 68.6% | 71.8% |
| Node + edge potential, no MI | 48.1% | 52.2% | 72.3% | 75.2% |
| Node + edge potential with MI | 49.2% | 53.5% | 74.2% | 77.8% |
| **Alignment recall on proteins with at least 256 non-redundant sequence homologs** | | | | |
| | 391 pairs in Set3.6K | | 509 pairs in Set2.6K | |
| Only with node potential | 59.5% | 63.4% | 71.3% | 75.8% |
| Node + edge potential, no MI | 62.1% | 66.7% | 73.5% | 78.1% |
| Node + edge potential with MI | 65.2% | 69.8% | 76.6% | 81.0% |

**Table 12.** Fold recognition rate of our method on SCOP40, with respect to the similarity (measured by E-value) between the test data and the training data.

| | E-value < 1e-35 | | | 1e-35 < E-value < 1e-2 | | | E-value > 1e-2 | | |
|---|---|---|---|---|---|---|---|---|---|
| | Top1 | Top5 | Top10 | Top1 | Top5 | Top10 | Top1 | Top5 | Top10 |
| hmmscan | 5.0% | 5.6% | 5.6% | 7.3% | 7.9% | 7.9% | 6.4% | 7.3% | 7.4% |
| | 10.3% | 14.5% | 15.8% | 9.7% | 12.9% | 13.5% | 11.6% | 16.5% | 17.5% |
| HHsearch | 16.0% | 23.2% | 26.5% | 18.5% | 26.2% | 30.3% | 18.9% | 27.2% | 31.7% |
| HHblits | 16.9% | 23.1% | 25.5% | 20.8% | 27.4% | 28.9% | 20.2% | 28.3% | 31.1% |
| MRFalign | **25.5%** | **35.9%** | **39.4%** | **29.7%** | **39.5%** | **43.3%** | **29.4%** | **39.0%** | **43.6%** |